# SideRand: A Heuristic and Prototype of a Side-Channel-Based Cryptographically Secure Random Seeder Designed to Be Platform- and Architecture-Agnostic

JV ROIG, Advanced Research Center – Asia Pacific College

Generating secure random numbers is vital to the security and privacy infrastructures we rely on today. Having a computer system generate a secure random number is not a trivial problem due to the deterministic nature of computer systems. Servers commonly deal with this problem through hardware-based random number generators, which can come in the form of expansion cards, dongles, or integrated into the CPU itself. With the explosion of network- and internet-connected devices, however, the problem of cryptography is no longer a server-centric problem; even small devices need a reliable source of randomness for cryptographic operations – for example, network devices and appliances like routers, switches and access points, as well as various Internet-of-Things (IoT) devices for security and remote management. This paper proposes a software solution based on side-channel measurements as a source of high-quality entropy (nicknamed "SideRand"), that can theoretically be applied to most platforms (large servers, appliances, even maker boards like RaspberryPi or Arduino), and generates a seed for a regular CSPRNG to enable proper cryptographic operations for security and privacy. This paper also proposes two criteria – *openness* and *auditability* – as essential requirements for confidence in any random generator for cryptographic use, and discusses how SideRand meets the two criteria (and how most hardware devices do not).

## 1 INTRODUCTION

### 1.1 Generating Random Numbers for Privacy and Security

The ability to generate strong random numbers is essential to cryptography, and central to security and privacy in the IT world. For our encryption technologies to function as expected, we rely on cryptographically secure pseudo-random number generators (CSPRNG) to actually produce high-quality random numbers suitable for cryptography. (For the purposes of this paper, "secure random number" will be used as shorthand to refer to "high-quality random number suitable for cryptography"; this refers mostly to how the random number was derived, and not any physical or inherent characteristic of any particular number itself)

Having a computer system generate a secure random number is a difficult task due to the deterministic nature of computers. Although random-generating algorithms have long existed, such as linear congruential generators (LCG) or the Mersenne Twister, these are not CSPRNGs (i.e, not suitable for use in cryptography). Hardware-based random number generators (called TRNGs or True Random Number Generators) are a common solution, particularly for servers that require massive amounts of entropy. TRNGs typically measure quantum random properties such as nuclear decay, or classical random properties such as thermal noise or atmospheric noise.

Being essential to security, some TRNGs have already made their way into CPUs themselves. VIA C3, released in 2003, has a TRNG built-in marketed as the VIA Padlock RNG [7, 22]. Intel also baked-in a TRNG into their CPUs starting in 2011 with the release of the Ivy Bridge architecture. [15, 17]

### 1.2 Appliances and Devices

While traditional servers may consider the problem of generating secure random numbers solved due to easy access to TRNGs (a point I dispute in the next section), the world's security infrastructure does not rest solely in the hands of these servers. Heninger et al [18, 19] found widespread factorable and duplicate TLS and SSH keys due to embedded devices suffering from "boot-time entropy hole". This reveals a problem, mostly an economic/financial one, that also has to be solved: cheap devices and appliances do not have integrated TRNGs in them, and the resulting lack of entropy has caused a failure in their cryptographic protocols, which ended up producing thousands of duplicates of keys for TLS certificates in the wild and vulnerable RSA and DSA keys.

Ideally, manufacturers or vendors should shell out the money to make sure their devices deal with entropy (or its shortage), equipping their devices or appliances with a TRNG somehow. In the real world, pragmatism tends to miss out, and if the added cost of including a TRNG does not make financial sense, manufacturers and vendors will continue selling vulnerable devices.

A software solution here would be superior, especially if one can be applied cost-free (such as Open Source software) and the implementation is also simple and straightforward (i.e., not a significant burden to their existing development team)

**1.3 The problem with hardware-based TRNGs**
In late 2013, due to the Snowden revelations, the TRNGs that were integrated into the CPUs themselves – Intel's DRNG and VIA's Padlock RNG – have fallen into suspicion [11, 12].

Whether the NSA has truly backdoored these by compelling Intel / VIA is not the central problem. The real problem is that these implementations are effectively blackboxes and are impossible to audit, especially in a live environment (there's no way you can actually audit the hardware circuit without destroying your CPU). For something so essential and central to cryptography and our security and privacy, we should not be depending on blackboxes.

Hardware devices also eventually fail. Aside from possibly being open and auditable, an acceptable software solution is superior to TRNGs in this regard, since software does not go bad like hardware does. Hardware random number generators do come with safety and health checks, but this is not a total safeguard – TRNG failure can result in service interruption that lasts until the specific device is replaced, since the computer system relying on it may have no other source of secure random numbers.

**1.4 Proposed Criteria for Confidence in Sources of Secure Random Numbers**
I propose two important criteria for sources of secure random numbers:

1. *Openness* - being so central to privacy and security, whatever our source of secure random numbers should be truly open, as in Open Source – with the entire algorithm available – and not a blackbox. If the algorithm is truly a CSPRNG, then the algorithm itself need not be secret. The method of generating secure random numbers should be easily studied and reviewed by interested parties. This is essential so that the very core of our security and privacy technologies is not something that can easily be backdoored or otherwise tampered with by nation-states and their security agencies.

2. *Auditability* – an extension of Openness, this criteria demands that aside from being open in spirit (which is easily accomplished by releasing something under an Open Source license and hosting the code publicly, such as through GitHub), the program itself is *practically* auditable – that is, it is as simple as possible, and avoids forms of programming or implementation "cleverness" that make the act of review harder.

These criteria are not meant to judge the quality of sources of random numbers – the quality of a PRNG and its suitability for cryptographic use is already well-served with entropy estimations (which I discuss in the Analysis section) along with necessary properties. Instead, these criteria deal with our confidence in the source, i.e., the level of trust – e.g., confidence that it has not been tampered with, backdoored, or otherwise obviously broken in some manner.

A 25,000-line algorithm that is mostly monolithic, for example, can easily be released as Open Source and hosted in a public GitHub repository. However, while it passes criteria #1 easily, it may not pass criteria #2, since being a monolithic, 25,000-line monster will make it extremely hard or impractical to thoroughly review and audit, which will lessen confidence in it as a trusted source, regardless of the quality of its output. In such a case, we would be better off with a similarly open, but far less complex implementation, which is easier to have reviewed and audited, and therefore more trusted.

**1.5 Roadmap**
The rest of the paper deals with the author's development of SideRand – a heuristic and prototype (coded in Python 3) that produces randomness based on side-channel measurements (solving the "computers are deterministic" wall that prevents CPUs from casually producing secure random numbers). Throughout the development of SideRand, including the incubation stage where ideas were considered, refined, and discarded, I was guided by the two proposed criteria for confidence in sources of secure random numbers. In particular, criteria #2, auditability, has resulted in the incremental prototypes of SideRand becoming more and more streamlined as possible, and removing instances of "cleverness" and replacing them with more obvious, straightforward code.

**2. BACKGROUND**

**2.1 Shortcomings of the Current State of Secure Random Number Generation**
Being essential to cryptography and central to our security and privacy, the generation of secure random numbers was a topic that interested me. Over the Christmas holidays of 2017, I thought about this problem, looked at the existing alternatives, and decided that the current state is severely lacking and needs to be improved. Specifically:

1. *Reliance on hardware-based TRNGs must end or be mitigated* – hardware devices effectively act as blackboxes, are impractical to audit in general, and next to impossible to audit within a live environment. Due to their blackbox nature, they are a natural target for nation-states and their intelligence apparatus. Whether the NSA / China / Russia has or has not backdoored

(through trickery or coercion) any TRNG is not the real problem – the problem is that they have been given a target to backdoor due to widespread reliance on hardware random number generators that are next to impossible to audit.

2. *Current software solutions are lacking* – software solutions either do not have enough theoretical backing, or are so complex that it makes them impractical to review and audit, and thus prone to suspicion; sometimes, both are true [2, 3, 4, 14]. There is also no solution currently available that is purposely designed as an architecture-and-platform-agnostic heuristic.

**2.2 Timing Variability – the Benchmarker's Bane**
Variability of benchmark runtimes has long been a bane for hardware reviewers, testers, researchers, developers, and most other users that rely on benchmark performance for key decisions (for example, whether a particular code change has actually sped up or slowed down a particular function). When measured with enough precision, benchmark runtimes can vary wildly, and are generally irreproducible. This applies to CPU benchmarks, GPU benchmarks, benchmarks of other hardware (hard disks, SSDs, etc), and especially to benchmarks that combine many of these components.

This was where I first started to imagine what would end up as the basis of the SideRand prototype – can CPU benchmark variability be used as the basis of entropy collection to generate secure random numbers? This presents itself as an interesting target for research and testing, since depending on variance of a benchmark runtime means using a side-channel measurement, instead of the actual value of any mathematical operation which would be deterministic. If made to work, relying on a side-channel measurement of a CPU would go a long way to solve the 2 problems mentioned in the previous section:

1. Reliance on hardware TRNG will be removed or mitigated, since everything that has a CPU – from large servers to small embedded devices or appliances – can potentially benefit.
2. This type of software solution, being based on a side-channel, may survive cryptanalysis, since it does not depend on an algorithm that may produce cyclical output.

**2.3 CPU Variance**
Assuming for now that the timing variability can be made to collect enough entropy to be suitable for cryptography, one obvious shortcoming that needs to be addressed is the "same-CPU" weakness. That is, if one specific CPU (say, an Intel i7-7700K) produces 100,000 different unique runtimes for a specific benchmark (with the variations being measured in nanoseconds), can an attacker produce the same 100,000 unique runtimes (thereby potentially making the proposed side-channel-based RNG predictable) if she buys the exact same CPU and runs the exact benchmark? This means a potential fatal weakness would simply be: *"find out potential possible CPUs running in the target's data center, buy these and make a table of potential values"*

Fortunately, the answer here is: CPU performance varies, even between two CPUs of the exact model, family and stepping. Researchers from Lawrence Livermore National Laboratory, for example, published a 2017 paper detailing an empirical survey of the variation in performance and energy efficiency in their clusters of servers [5]. After characterizing the performance and energy efficiency of 4,000 CPUs, they found that no two processors had identical performance characteristics. This variation has not been improving (i.e., not becoming less pronounced) as CPUs become more modern; instead, from Sandy Bridge (2$^{nd}$ generation Intel Core architecture) to Ivy Bridge (3$^{rd}$ generation) to Broadwell (5$^{th}$ generation), the variation in performance has increased between processors of the same model, family and stepping. I also present empirical data specific to SideRand that reinforces this, later on in the Entropy Analysis section.

In a nutshell, since no two CPUs perform identically (given enough precision in measurement), relying on a side-channel measurement based on benchmark runtime is not trivially exploitable by merely purchasing the same CPU model.

**3. DESIGN AND EVOLUTION OF SIDERAND**

**3.1 Fundamental Design Notes**
I developed SideRand based on the variability of benchmarks, and the design of which is guided by my proposed criteria for confidence in sources of secure random numbers (*openness* and *auditability*). This section documents the evolution of SideRand prototypes, from the first attempt (designated as "mark 1" or "mk1"), until the final prototype version (currently, Siderand mk10). I've opted to discuss the evolution of the design from mark 1 to mark 10, instead of discussing only the final design of the mark 10 prototype, because walking through the evolution of the prototype serves to better illustrate the problems and pitfalls that would be encountered in creating a CSPRNG based on CPU side-channel measurements. Just as well, walking through the evolution of the prototypes will serve to answer common questions that may arise if only the mk10 final prototype was discussed, especially about design decisions and tradeoffs.

During the discussion of the specific versions of the SideRand prototype, some parts of the code will be displayed. However, the full Python 3 source codes of each are also available online through the author's SideRand site: http://research.jvroig.com/siderand. It is recommended that the reader refer to the actual, complete source codes for clarity. If the reader also has a Python 3 interpreter installed, the prototypes are executable and can be tested with minimal effort.

**3.2 SideRand mk1**
The mark 1 prototype functioned as follows:

- A set of 100 random numbers is used as reference (this set of random numbers was generated using */dev/urandom* of Linux) and imported as *random_set*.
- Part 1: Each of the 100 random numbers is matched against each other, executing various math operations. Time to execute a set of operations is recorded into an array that ends up with 100 runtimes, called *tot* ("total times", set of all runtimes).
- Part 2: This array is then processed so that 50 differences (diffs) between runtimes are produced – the last (100th) runtime is subtracted from the first runtime to get their difference, then the 99$^{th}$ runtime is subtracted from the 2$^{nd}$, and so on. Each diff is then hashed using SHA256. Effectively, the diffs capture the variances between two runtimes.

Here, "Part 1" and "Part 2" are merely logical distinctions to make discussions easier, particularly across prototype versions. Part 1 runs the benchmarks and collects the runtime measurements. Part 2 produces the diffs from the measurements collected, and generates the output based on the diffs through SHA256.

The source code for SideRand is shown below.

**ALGORITHM 1:** SideRand mk1

```
import time
from random100 import random_set
import hashlib

def get_entropy():
    limit = len(random_set) - 1

    scale = 50
    tot = []
    i = 0
    while i <= limit:
        j = 0
        while j <= limit:
            k = 0
            time_s = time.time()
            while k < scale:
                a1 = random_set[i] + random_set[j]
                a2 = (random_set[i] - random_set[j])
                a3 = (random_set[i] * random_set[j])
                a4 = (random_set[i] / random_set[j])
                a5 = (a3 * random_set[i]) / (a4 * random_set[j])
                k += 1
            time_e = time.time()
            tot.append((time_e - time_s))
            j += 1
        i += 1

    f = open('random_numbers','wb')
    limit = int(len(tot) / 2)
    i = 0
    while i < limit:
        diff = tot[i] - tot[limit - i]
        if diff > 0:
            h = hashlib.sha256(str(diff).encode())
            f.write(h.digest())
        i += 1

get_entropy()
```

Algorithm one shows the entire source code, from imports, subroutine definition (including the "def" call itself), and the final line that calls the subroutine. Since only the subroutine definition itself changes, most of the succeeding Algorithm sections will show only the subroutine definition or specific portions of it, omitting the import lines, *def* line, and the *get_entropy()* line.

There is some good here already, but a lot more bad.

The good part is simply that the code is very short, essentially a nested-loop in order to match every element of the set of the 100 random numbers to each other. If it works as expected, then it would be easy to audit due to length and simplicity.

The bad part is that it does not actually work – even though the output is a SHA256 hash, it still fails the FIPS 140-2 statistical test, as implemented in Linux through the *rngtest* module. This shows that it is horribly broken.

### 3.3 SideRand mk2
The mk2 prototype tries to fix mk1 by making a change in the implementation of Part 2, shown below.

**ALGORITHM 2:** SideRand mk2 source code, Part 2 area only

```
limit = int(len(tot)) #use if diff = tot[i] - tot[i + 1]
i = 0
while i < limit:
    diff1 = (tot[i] - tot[i + 1])
    diff2 = (tot[i] * tot[i + 1])
    diff3 = abs(tot[i] - tot[i + 1])
    diff = str(diff1) + str(diff2) + str(diff3)

    h = hashlib.sha256(str(diff).encode())
    h.update(str(tot[i]).encode())
    h.update(str(tot[i+1]).encode())
    f.write(h.digest())

    i += 2
```

The change here made Part 2 more complicated. Instead of a simple subtraction operator to produce the diffs, it is now a combination of subtraction, multiplication, and absolute value subtraction, all of which are then concatenated together.

There is still not much good to be had here. In fact, this attempt is arguably worse than mk1: it has become more complicated, but still can only pass *rngtest* below 90% of the time. It is still horribly broken.

**3.4 SideRand mk3**

The mk3 prototype made another change to Part 2, shown in Algorithm 3. This time, an attempt to clean the measurements is made by removing all leading zeroes, decimal points, and spaces from each diff – for example, a measurement of "0.000000012345" will now just be "12345". The hashing portion is also changed – instead of immediately hashing one diff, the entire set of diffs are hashed. The way the diffs are generated have also been changed, and are now separated into batches.

The changes here have made the SideRand mk3 prototype pass ~95% of the type – a marked improvement, but still well below what is expected from a CSPRNG.

What should the passing rate be? The passing rate should be ~99.92%, which is what the Linux random devices (*/dev/random* and */dev/urandom/*) obtain when they are piped to *rngtest*. I obtained this figure from a different experiment (2017, still to be published) that passed over 40 terabytes of randomness from */dev/random* and */dev/urandom* to *rngtest*. Since that experiment cannot be cited yet as it is also still being prepared for publication, I'm sharing a simple way to verify the ~99.92% figure. On any Linux system with *rngtest* available (courtesy of the *rng-tools* package) simply run the following:

    rngtest -c 5000 < /dev/urandom

In any modern Intel processor, you can also test */dev/random* the same way:

    rngtest -c 5000 < /dev/random

Tabulating multiple runs will converge to ~99.92% passing rate for both. You may also look at the aforementioned experiment data through the website for that experiment: http://research.jvroig.com/linuxrand

Since 95% is still well below the ~99.92% target, SideRand mk3 is still broken. Worse, the code has become progressively more complicated, shown below.

**ALGORITHM 3:** SideRand mk3 source code, Part 2 area only

```
batch = 10
limit = int(len(tot)) / batch
i = 0

megadiff = []
while i < limit:
    diff = []
    j = 0
    while  j < batch:
        j += 1
        string_num  = "{:22.21f}".format(abs(tot[i] - tot[i+j])).replace(".","").replace(" ","")

        # find where the 0s stop
        k = 0
        mark = 0
```

```
        length = len(string_num)
        while k < length:
            if string_num[k] != '0':
                mark = k
                break
            k += 1

        if k < length:
            string_num = string_num[mark:length]
            diff.append(string_num)
            megadiff.append(string_num)

        h = hashlib.sha256(str(set(diff)).encode())
        f.write(h.digest())
    i += 10
```

### 3.5 SideRand mk0

As a sanity check, a mk0 prototype was created, just to verify that something must indeed be terribly broken in all previous prototypes since even a SHA256-hashed output fails to pass *rngtest* sufficiently. The mk0 prototype is nothing but an incrementor, from 0 to 100,000 with each step (in increments of 1) being hashed through SHA256, as displayed in Algorithm 4.

**ALGORITHM 4:** SideRand mk0 – not a real prototype, but a sanity-check

```
limit = 100000
with open('random_numbers','ab') as f:
    i = 0
    h = hashlib.sha256(str(i).encode())
    while i < limit:
        i += 1
        h.update(str(i).encode())
        f.write(h.digest())
```

The goal of this sanity check was to verify that the fundamentals are correct with regards to SHA256 – if it is not asked to hash repeating inputs, then its output should pass *rngtest* with a rate that is no different than output from */dev/urandom* or */dev/random*. And indeed this was the case. SideRand mk0 easily passes ~99.92% of the time. This small experiment shows that all previous prototypes must have ended up supplying values that repeat enough that even a SHA256 output fails *rngtest*.

### 3.6 SideRand mk4

The mk4 prototype addresses the weaknesses of the first 3 prototypes by collecting more samples (i.e., more timing info due to having much more operations), creating a major change in the Part 1 area of the source code. Significant change in the Part 2 area was also done, particularly for the introduction of the "stretch" parameter, which applies key stretching in order to produce more bits in the output. Key stretching does not affect the actual entropy of the system, merely its suitability for practical usage (i.e., increasing the throughput of random bits creation, not increasing the system entropy).

SideRand mk4 is the first actual prototype that passes *rngtest* with the same passing rate as the Linux random devices - ~99.92%.

While this is good, the code has become cluttered, complex, and is marked with "cleverness" - for example, tons of operations are done because they seem random, chaotic and unpredictable. Unfortunately, there is really no justification for them from a theoretic perspective – for example, why those particular operations are repeated and recorded twice, and why the fifth set of operations (a combo of multiplication and division) was designed as it was. I believe this does not reflect the ideals that the two criteria for confidence in sources of secure random numbers strive for – not only is there no theoretical backing to these complex operations, they are nothing more than "cleverness" that hinder auditability.

**ALGORITHM 5:** SideRand mk4 source code for the Part 1 area. It has become longer for no justifiable reason other than it seems to work

```
limit = len(random_set) - 1

tot = []
i = 0
while i <= limit:
    j = 0
    while j <= limit:
        time_s = time.time()
        a1 = random_set[i] + random_set[j]
        time_e = time.time()
        tot.append((time_e - time_s))
```

```
            time_s = time.time()
            a1 = random_set[i] + random_set[j]
            time_e = time.time()
            tot.append((time_e - time_s))

            time_s = time.time()
            a2 = (random_set[i] - random_set[j])
            time_e = time.time()
            tot.append((time_e - time_s))

            time_s = time.time()
            a2 = (random_set[i] - random_set[j])
            time_e = time.time()
            tot.append((time_e - time_s))

            time_s = time.time()
            a3 = (random_set[i] * random_set[j])
            time_e = time.time()
            tot.append((time_e - time_s))

            time_s = time.time()
            a3 = (random_set[i] * random_set[j])
            time_e = time.time()
            tot.append((time_e - time_s))

            time_s = time.time()
            a4 = (random_set[i] / random_set[j])
            time_e = time.time()
            tot.append((time_e - time_s))

            time_s = time.time()
            a4 = (random_set[i] / random_set[j])
            time_e = time.time()
            tot.append((time_e - time_s))

            time_s = time.time()
            a5 = (a3 * random_set[i]) / (a4 * random_set[j])
            time_e = time.time()
            tot.append((time_e - time_s))

            time_s = time.time()
            a5 = (a3 * random_set[i]) / (a4 * random_set[j])
            time_e = time.time()
            tot.append((time_e - time_s))

            j += 1
        i += 1
```

### 3.7 SideRand mk5

The mk5 prototype is an attempt to streamline the mk4 prototype, while retaining the *rngtest* passing rate. In the Part 1 area of the code, the main loop has been significantly simplified and now only does 1 operation. In the Part 2 area of the code, much of the simplicity was achieved by removing the codes that removed zeroes, decimal points, and formatting. These were all unnecessary, since all values would be run through SHA256 anyway, and were not the source of failures of earlier prototypes (rather, they failed due to duplicate diffs, which formatting and removal of leading zeroes would not mitigate in the slightest).

Also in the Part 2 area of the code, creating the diffs, which used to be a subtraction operation (hence, the "diff" term), was changed to a multiplication operation. Tests showed that the diffs gathered had more unique values when the collected runtimes were multiplied instead of subtracted. Intuitively, this also checks out, since a subtraction operation in a set of numbers are more likely to result in similar values. For example, a set of runtime values containing 0.0003, 0.0002, and 0.0001, when each element is subtracted with each other, we would get the following values: 0.0001 (from 0.0003 – 0.0002), .0002 (0.0003 – 0.0001), 0.0001 (0.0002 – 0.0001). With the same set and distribution, but this time using a multiplication operation, we would get no duplicate values: 0.00000006, 0.00000003, and 0.00000002. For our case, then, subtraction lessens the entropy we can collect, whereas multiplication does not – or, at least, at this stage in the prototypes, lessens the entropy much less than a subtraction operation would.

Algorithm 6 shows the entirety of SideRand mk5 code, now back to being more manageable thanks to the streamlining of mk4 code, with only about 50 lines of code.

**ALGORITHM 6:** SideRand mk5 source code – much simpler than mk4

```
    limit = len(random_set) - 1
```

```
        tot = []
        i = 0
        while i <= limit:
            j = 0
            while j <= limit:
                samples = 10
                k = 0
                while k < samples:
                    time_s = time.time()
                    a1 = random_set[i] + random_set[j]
                    time_e = time.time()
                    tot.append((time_e - time_s))
                    k += 1
                j += 1
            i += 1

        f = open('random_numbers','ab')
        limit = len(tot)
        i = 0
        diff = []
        while i < limit:
            num = tot[i] * tot[i+1]
            diff.append(num)
            i += 2

        stretch = 10000
        i = 0
        time_s = time.time()
        h = hashlib.sha256(str(diff).encode())
        digest = h.digest()
        f.write(digest)
        time_e = time.time()
        time_p = time_e - time_s

        while i < stretch:
            time_s = time.time()
            i += 1
            h.update(digest + str(i).encode()
                            + str(time_s).encode()
                            + str(time_e).encode()
                            + str(time_p).encode())
            digest = h.digest()
            time_e = time.time()
            time_p = time_e - time_s

            f.write(digest)
```

### 3.8 SideRand mk6

SideRand mk6 was an experiment to see if a smaller random set of numbers can be used as the basis for the operations in the Part 1 area (the benchmarking code). All previous prototypes used a set of 100 random numbers; mk6 only used a set of 20 random numbers as the basis of its benchmark operations. Since far less computations will be executed, the samples parameter was increased – from an original value of 10, new values of 100 and 250 were also tested.

The *rngtest* passing rate was unchanged, at still ~99.92%. However, with a samples parameter of only 10 (same as mk5), unique values produced per execution dropped – whereas mk5 would typically report unique values of 100-150 per run, mk6 would only have about 20-40 unique values. Increasing the samples parameter to 100 bumps up the typical unique values per run to ~70-90. Increasing samples to 250 results in typically the same number of unique values per run as mk5.

### 3.9 SideRand mk7

The mk7 prototype tested out different timer functions in Python3, specifically *time.time()* (which is what all previous prototypes have been using), *time.clock()*, *time.perf_counter()*, and *time.process_time()*. Of the 4, *time.clock()* was immediately disqualified, due to the official docs declaring it deprecated.

From testing, both *time.perf_counter()* and *time.process_time()* have at least 100x better resolution than *time.time()* in Linux, which makes them far more ideal for benchmarking. In Windows, *time.time()* has an extremely low resolution, 16ms, and isn't a viable timer. Having better resolution (more precision / decimal places for runtimes) directly gives the system more entropy overall, as otherwise unmeasurable timing differences using *time.time()* are now measurable. Unique values per run have skyrocketed from ~100-150, to about 5,000 per run, due to the significant increase in timer resolution. Timer resolution varies between platforms, but Python 3 makes it easy enough to check by simply using *time.get_clock_info()* [20].

The final choice for an improved timer function was *time.perf_counter()*, since testing revealed that *time.process_time()* does not wok as expected in Windows platforms. Official documentation on *time.process_time()* also mention that it tracks only the execution time of the process itself (hence the name), and ignores slowdowns caused by external factors. This is not optimal for SideRand usage, since any variance, internal or external to the process itself (such as task switching due to other background tasks), is useful to SideRand as more entropy.

### 3.10 SideRand mk8 and SideRand mk8b
There are still items to improve from the SideRand mk7 prototype:
1. It still uses a set of 20 random numbers (which it inherited from mk6)
2. Relying purely on *time.perf_counter()* to be supported may not be optimal.

The first concern is for the desire for simplicity and utmost *auditability*. While a set of 20 random numbers is much smaller than a set of 100 random numbers, it's still a set of random numbers that we import, and that according to the premise of SideRand (timing variance / runtime irreproducibility, with enough precison), shouldn't matter at all. To fix this, the mk8 prototype does away with using a set of random numbers. Instead, two values used are fixed. The exact values themselves are not important; the values that the mk8 prototype used are just two numbers generated at random, and then hard-coded into the prototype code. Figure 7 shows the Part 1 area of the mk8 prototype, with the hard-coded numbers used in the benchmark operation. Using just 2 fixed values (compared to 100 and 20 of the previous prototypes) resulted in no difference in *rngtest* passing rate.

**ALGORITHM 7:** SideRand mk8 source code for the Part 1 area – *val1* and *val2* are just numbers that were randomly generated to be placed in the code, with no special significance.

```
val1 = 2585566630
val2 = 576722363

tot = []
samples = 100
scale = 1
i = 0
while i < samples:
    time_s = time.perf_counter()
    j = 0
    while j < scale:
        a1 = val1 + val2
        j += 1
    time_e = time.perf_counter()
    tot.append((time_e - time_s))
    i += 1
```

The second concern, reliance on *time.perf_counter()*, is addressed by the addition of a scale factor (the *scale* variable). This is meant to be increased when the available timer does not have a high enough resolution. In mk8b, *time.perf_counter()* is swapped out for *time.time()*, to simulate a scenario where the running system (which may be a VM, or a low-end device with a small, cheap processor) has a lower resolution. With a much lower resolution timer, the benchmark workload has to be increased so it is more measurable, so the *scale* factor in mk8b was scaled up from 1 to 250. Scaling up the workload increased the unique values per run, demonstrating that the scale factor can be used to compensate for a lower-resolution timer. This does increase runtime (since that is exactly what *scale* is supposed to do – make the benchmark slower so that a lower-resolution timer can better distinguish one run from another), so there would be a practical limit to how low resolution the timer can be, also affected by how slow or fast the processor is.

**ALGORITHM 8:** SideRand mk8b source code for the Part 1 area – similar to mk8, with the scale variable set to 250, and the use of *time.time()* instead of *time.perf_counter()*, to simulate a lower-res timer being used with an increased scale.

```
val1 = 2585566630
val2 = 576722363

tot = []
samples = 100
scale = 250
i = 0
while i < samples:
    time_s = time.time()
    j = 0
    while j < scale:
        a1 = val1 + val2
        j += 1
    time_e = time.time()
    tot.append((time_e - time_s))
    i += 1
```

## 3.11 SideRand mk9

The mk9 prototype improves the mk8 prototype with a change in the Part 1 area of the code that is merely superficial, and a change in the Part 2 area of the code that affects the hash stretching done.

In the Part 1 area of the code, *tot* (which originally meant *total times*) was simply renamed to the more informative *times*. The stretch variable, used for key stretching the hash output in Part 2, was also moved up to Part 1 along with the other defined variables.

The changes in the Part 2 area improve the multiplication operation that creates the diffs (adding a 1 to each operand to make sure any zero readings on either side of the operands do not result in a zero result), and also streamline the key stretching operation, removing all unnecessary cruft. Previously, the hash stretching portion also had timer values (start time, end time, process time) and an incrementor, all of which are mixed in with the digest of the diffs. This was done to make key stretching safer – there's always new content in each iteration of the stretching. However, these are all unnecessary. If the collected entropy is truly impractical to predict or brute-force (which should be the case in our quest to generate secure random numbers), then adding more measurements just to key stretch is simply more work for no gain. And the reverse is also true – if in the first place the collected entropy is predictable or brute-forceable, then adding more junk into the key stretch operation doesn't make the collected entropy any better. The mk9 prototype then strips away all this junk, the key stretching operation is now just the previous digest mixed with the entire diff set, i.e., the actual, supposedly unpredictable and un-brute-forceable entropy.

**ALGORITHM 9:** SideRand mk9 source code

```
val1 = 2585566630
val2 = 576722363

times = []
samples = 100
scale = 250
stretch = 100

i = 0
while i < samples:
    time_s = time.perf_counter()
    j = 0
    while j < scale:
        a1 = val1 + val2
        j += 1
    time_e = time.perf_counter()
    times.append((time_e - time_s))
    i += 1

limit = len(times)
entropy = []
i = 0
while i < limit:
    num = (1 + times[i]) * (1 + times[i+1])
    entropy.append(num)
    i += 2

entropy_bytes = str(entropy).encode()
h = hashlib.sha256(entropy_bytes)
digest = h.digest()

f = open('random_numbers','ab')
f.write(digest)

i = 0
while i < stretch:
    h.update(digest + entropy_bytes)
    digest = h.digest()
    f.write(digest)
    i += 1
```

## 3.12 SideRand mk10

The final prototype, mk10, introduces a much needed improvement in the entropy collection logic of the Part 2 area. The mk9 prototype improved on the multiplication operation (multiplying two of the collected run times to each other) so that a zero value on one operand would not result in a 0 (which biases collected entropy to having zeroes more common than other values). Recall that in the SideRand mk5 prototype discussion, it was mentioned that subtraction lessens entropy, while multiplication does not (or at least only does so at a much lesser degree). I realized, however, that even this multiplication operation is very unfriendly to entropy: from 100 measurements taken (the default sample size in mk8 and mk9), the way entropy was collected into the diffs array ended up with having only 50 resulting values – effectively cutting in half the samples we took.

The real problem, of course, is that trying to subtract or multiply values to each other is a poor attempt at mixing. Not only did we end up halving the collected values (from 100 varied measurements, down to 50), we ended up doing entropy collection twice – once when we got 100 samples, then again when we derived 50 values from those 100 samples.

The best way to mix these values would be to just hash them together. The mk10 prototype was then written to remove the multiplication operation. Instead, all 100 collected samples are hashed as is. The resulting hash is then subjected to key stretching, as before. Since there is no more multiplication operation in-between the initial samples collection and the hash operation, we save about 8 lines of code and increase performance. Best of all, the entropy of the system is now more straightforward, which will help us estimate the amount of entropy this prototype (and the heuristic in general) can have. The entropy gathered is simply the runtime values the benchmark produces and the exact order they appear when we collect measurements.

**ALGORITHM 10:** SideRand mk10 source code

```
val1 = 2585566630
val2 = 576722363

times = []
samples = 100
scale = 250
stretch = 100

i = 0
while i < samples:
    time_s = time.perf_counter()
    j = 0
    while j < scale:
        a1 = val1 + val2
        j += 1
    time_e = time.perf_counter()
    times.append((time_e - time_s))
    i += 1

samples_concat = str(times)
diff_bytes = samples_concat.encode()
h = hashlib.sha256(diff_bytes)
digest = h.digest()

f = open('random_numbers','ab')
f.write(digest)

i = 0
while i < stretch:
    h.update(digest + diff_bytes)
    digest = h.digest()
    f.write(digest)
    i += 1
```

Now that a prototype exists that has a sufficiently simple and concise code, and that passes a rudimentary statistical analysis (which merely means it is not *obviously broken*, but proves nothing quality-wise from the perspective of generating secure random numbers), we can proceed with analyzing the amount of entropy the proposed heuristic and prototype.

**4. ENTROPY ANALYSIS**

**4.1 "The Entropy Source We Deserve, But Not the One We Need Right Now"**
The output of any random number generator cannot be used directly as proof of its suitability as a source for secure random numbers. Fortunately, our cryptographic protocols also do not require that the source is true randomness. Instead, cryptographically secure random number generators merely need to be sufficiently unpredictable and have a ridiculously large key space (set of possible outputs) such that brute-force attacks are infeasible within the applicable threat-model. To paraphrase Gordon in "The Dark Knight (2008)", "*True randomness is the entropy source we deserve, but not the one we need right now.*"

The proposed heuristic, as implemented in the SideRand mk10 prototype, will be analyzed to see if it functions with enough unpredictability and with a large enough key space to make it a reliable source for secure random numbers. Throughout the tests, the scale factor within the SideRand code is tweaked to be appropriate to the speed and timer resolution of the machine. In general, the scale factor has a direct relationship to CPU speed, and an inverse relationship to timer resolution – the faster the CPU, the greater the scale factor, whereas the greater the timer resolution, the lesser scale factor is needed. The worst case would be a very fast CPU, with a very low-res timer, such as using *time.time()* in Python 3 under Windows.

**4.2 Key Space – Part One**
To get data to estimate the key space involved, a "unique_logger" tool was designed. This tool has two parts:

1. A file that is a slightly modified form of the SideRand prototype – instead of hashing the runtimes it collects, it logs them all into a text file, one runtime per line.
2. A separate Python script that parses the resulting log file to catalog the number of unique values generated.

Several different machines were tested using the unique_logger tool running SideRand mk10. The raw results, as well as the tool itself for the reader's own testing or inspection, can be found in the SideRand website linked to in section 3. In summary, the key space analysis shows that the runtimes measured can have thousands of unique values, which depend largely on CPU speed, available timer resolution, and the software stack.

With thousands of potential values - one of the best cases being 142,703 unique values logged - and with the raw random output (i.e., the value that will be hashed) being a chronological sequence of 100 of these values (i.e., the specific order they were collected in, not sorted in any way), that gives a ridiculous upper limit of ~ 140,000 ^ 100.

This is the potential upper bound. It's the stuff that the wild imaginations of a crypto-nerd (such as yours truly) are made of. However, this potential key space will be greatly affected by predictability – it doesn't matter if there is an unimaginably large set of possible values if, in practice, there are really only a handful of values that appear 99% of the time. To determine this, we need to analyze the distribution of the runtime values.

**4.3 Distribution**
To get data about the distribution of runtime values, another tool was created, and was still rather-unimaginably named: distribution_checker. The distribution_checker tool is composed of the following components:
1. A file that is a slightly modified form of the SideRand prototype – instead of hashing the runtimes it collects, it returns the entire set to the next component of the tool
2. The next component is the largest and most complex part of the tool (diff_distribution.py). This calls the first component, receives the set of runtime values, then logs them into an sqlite3 database. This also supports multithreading so that distribution tests can be done with varied amounts threads.
3. A set of bash scripts were also created to automate tests using the pair of Python 3 files.

As with the earlier "unique_logger" tests, several different machines were tested. Each machine ran the SideRand mk10 prototype 3,000 times, producing 300,000 runtime measurements per machine. The raw results, corresponding image files of the distribution graphs, as well as the distribution_checker tool itself, can be found as supplementary information in the SideRand website linked to in section 3. Note that the graphs are generally very high-resolution, composed of thousands of values along the x-axis, and cannot be faithfully reproduced in low-res within this paper without sacrificing the nuances that affect entropy estimation. Hence, the majority of the graphs are saved in the supplemental website and not reproduced here, except for a select few to serve as illustration.

In summary, the distribution found in tests with distribution_checker has the following characteristics:
1. The distribution is multi-modal
2. The distribution is not uniformly-random across the entirety of all unique values logged
3. The shape of the distribution varies, even between CPU's of the exact same model, family and stepping, running on the same OS and Python version – for example, one machine can show three clear high peaks with three clear lower peaks, while an identical machine shows something that is closer to a bimodal distribution (Figures 1 - 3).
4. If we limit the graph to a very small set of the most frequent values (e.g., considering only the top 20 most frequent values out of thousands of collected values), the distribution tends toward flatness, i.e., almost uniformly-random (Figure 4)

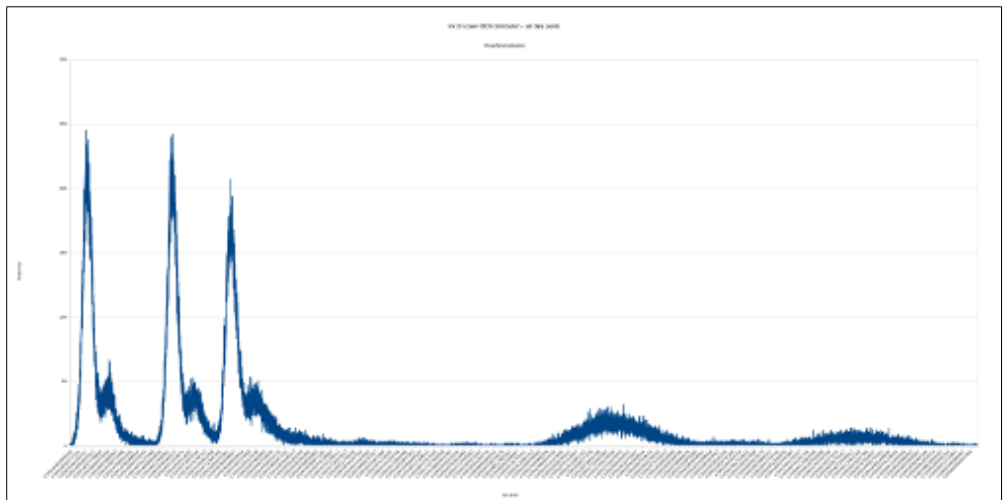
Figure 1: Frequency distribution of all values, "Constitution" server (21,509 unique values)

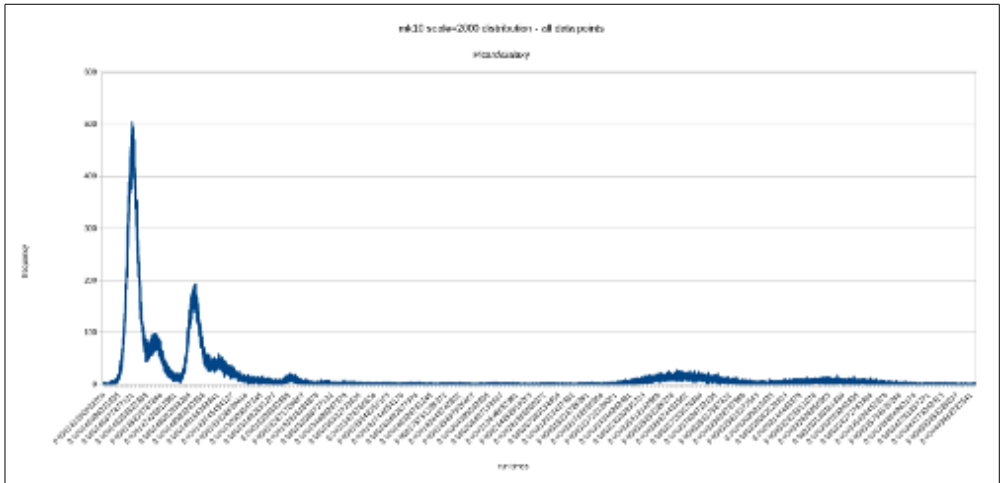

Figure 2: Frequency distribution of all values, "Galaxy" server (19,176 unique values)

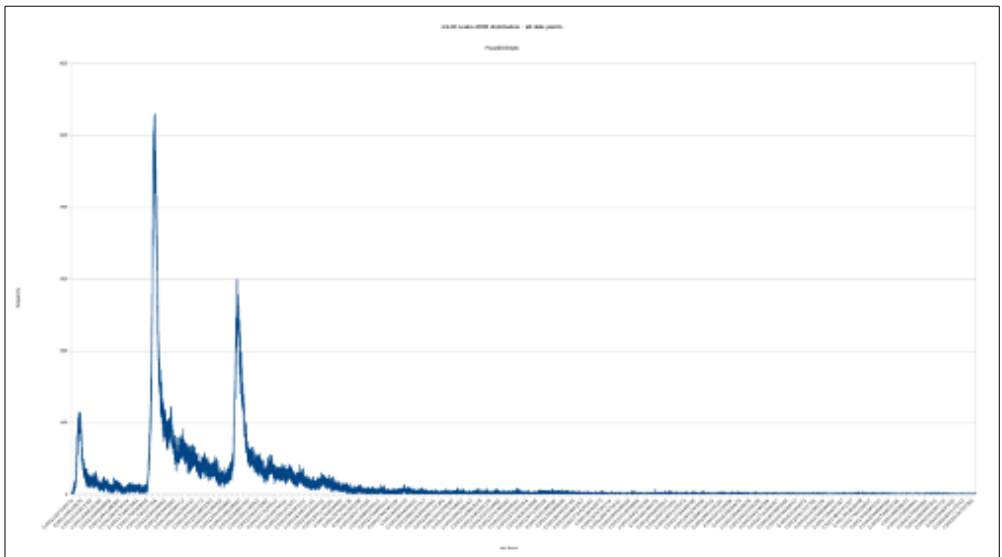

Figure 3: Frequency distribution of all values, "Intrepid" server (19,517 unique values)

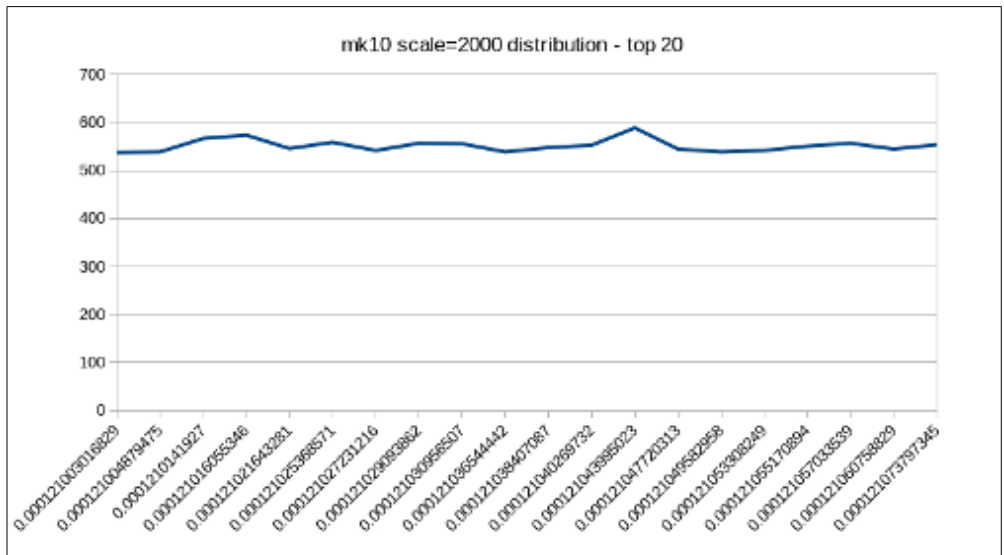
Figure 4: Frequency distribution of top 20 values, "Sovereign" server (out of 14,127 total unique values)

The graphs reproduced here are meant only to give a sense of the comparative differences in general shape (Figures 1 – 3) as well as the flat distribution of a small subset using the most frequent 20 values (Figure 4). Higher-resolution versions of these graphs for readability and better analysis can be found as supplemental information in the SideRand website linked to in section 3.

**4.4 Key Space – Part Two**
Before proceeding to analyze the entropy using the knowledge gained from the unique_logger and distribution_checker tools, revisiting key space estimation is in order, to note that there is far more variability here than we can realistically model. For example, while the graphs produced from the distribution_checker tests were compared and presented (found in supplemental information through the SideRand website), the actual values across the x-axis (runtimes) are not the same for every graph. The values are continuous in nature (nanoseconds or less). Comparing the top values of each server reveals little to no overlap, even across different machines with the exact same CPU and OS. For example, in a cluster of 4 servers with identical hardware (CPU, motherboard, RAM, harddisk) and identical OS (Linux - Fedora 25), comparing the top 500 runtime values collected by each server resulted in only two overlaps. From a total of 2,000 runtime values, 1,998 were unique across all four servers.

What this tells us is that, whatever we can estimate from any single test from a single machine, it is very likely that the theoretical entropy in this heuristic is actually far greater. For example, in the earlier key space estimation subsection, the example of one machine producing approximately 140,000 unique values was used to get a theoretical upper limit of $140,000^{100}$. But since these unique values, the specific runtimes recorded, are not the same for every machine (even among machines that have the exact same CPU model/family/stepping), then the actual unique values that can be produced in the real world (with varying CPUs, platforms, and even varying SideRand tuning) are far higher. For platforms that have better than nanosecond resolution, there can be billions of possible runtime values across different machines with different operating systems, CPUs, and software stack. Stringing a hundred of these billions of possible values together, as SideRand does, likely has far more entropy than needed by any cryptographic use case.

**4.5 Modeling Entropy Using Worst-Case Estimates**
With the distribution being chaotic and multimodal, and with the x-axis values themselves being inconsistent across machines, it's impossible to exactly model the entropy found in SideRand.

Fortunately, there is a convenient way to estimate the entropy to determine suitability as a CSPRNG. We can simply model a worst-case estimate based on the data collected, construct the worst-case estimate such that we know that the actual entropy is sure to be far higher, and then compare this worst-case estimate against the standard that we need. If the worst-case scenario already meets or exceeds the standard, then exactly modeling the entropy in the system is unnecessary since it would already be known to meet or exceed the required standard.

For a cryptographically secure seed, a standard to meet is an entropy of at least 256 bits (approximately $1.15 \times 10^{77}$), from which can be derived an unlimited number of keys using deterministic cryptography [8, 9, 10]. 128 bits of entropy was suggested by the IETF in 2005 [13], but this paper will adopt a more conservative stance and aim for the stricter standard of 256 bits.

SideRand worst-case can be estimated as follows:
1. Instead of up to hundreds of thousands of unique values, let us assume there are only 20 values. In most of our tests, this means we are effectively considering only the top 5% or less, conveniently ignoring at least 95% of the runtime values collected in order to model our worst-case estimate.
2. Based on distribution tests, the top 20 values are flattish, which we can treat as a uniformly random distribution.
3. SideRand collects 100 samples from those 20 possible values. With a uniformly random distribution, the key space is $20^{100}$.

This gives us a worst-case entropy estimate of 1.26x10^130 (20^100), which far exceeds the requirement of 256 bits of entropy (1.15x10^77). In reality, since we ignored around 95% of actual real values, we can be confident that the actual lower-bound of entropy in this system is far greater than 1.26x10^130. Since this already exceeds the standard of 256 bits of entropy for secure random numbers, then it is unnecessary to determine the exact lower bound for now. Whatever it actually is, it is already far more than needed for our cryptographic protocols.

### 4.6 Throughput Measurements
How long does one run of SideRand take?

In standard Intel or AMD-based machines running mainstream Linux distributions, SideRand with a scale of 2000 takes less than a tenth of a second on average. Specifically tested Linux distributions were Fedora 25, CentOS 7.4, and Ubuntu 16.04, and Ubuntu 17.10. Even "small core" architectures – such as Intel Atom and AMD Bobcat cores – take less than 0.1 seconds on average. Benchmark results are available as supplementary information in the SideRand website linked to in section 3.

This means solving the boot-time entropy hole problem for headless Linux servers, as described by Heninger et al in [18, 19] is inexpensive to solve. There is virtually no cost in terms of additional processing time to collect entropy that we estimated, at worst, to be well in excess of 256 bits.

### 4.7 Where the Variance Comes From
It may be worthwhile at this point to mention why this runtime variance between benchmark runs exists.

Modern CPUs contain millions to billions of transistors – even the ARM Cortex-A9, released over ten years ago (2007), has an estimated 26 million transistors. These transistors that make up a CPU are not perfectly uniform (as truly nothing humans create ever are, when measured with enough precision), and transistor variability has long been something that CPU designers cope with, such as designing for the worst case (guard-banding). Aside from transistor variability itself, there are also systematic variability (caused by manufacturing, with CPU binning as a common coping strategy) and local variability (random dopant fluctuation) [21]. These factors make CPUs unique from each other, even those that come from the same wafer and binned as the exact same model, family and stepping. CPU enthusiasts, especially overclockers, refer to this as the "silicon lottery".

This variation between CPUs that are sold as identical is not getting better (smaller). As the LLNL team found in [5], this variance has only increased with more recent processors (e.g., Broadwell microarchitecture compared to Ivy Bridge microarchitecture). The reason for this is the improvement in dynamic frequency scaling features in most CPUs – whereas these "turbo" features in multi-core architectures used to be very blunt (a fixed frequency if only 1 core is operating, a slightly lower fixed frequency with an additional core operating, etc), current implementations from Intel, AMD and ARM-based processors offer turbo-like features with smarter capabilities that take into account estimated current consumption, estimated power consumption, and processor temperature (respectively, Intel Turbo Boost Techonology 2.0, AMD SenseMI / Extended Frequency Range, ARM DVFS technology) [1, 6, 16] .

These variances that stem from the transistor level also affect repeated runs of the same physical CPU. Execution of the same instruction and data won't take the exact same path each time – i.e., each run would not be using the exact same transistors (be they transistors in the CPU registers, cache, or execution units). The physical location of the data, for example, also results in runtime variance due to latency differences – one run, for example, may have the data on physically one end of the cache, whereas the other run may have the same data on the other end of the cache. These variations can be as small as nanoseconds or fractions of nanoseconds.

Aside from the practical, real-world consideration of needing a CPU and platform that offers reasonable timer resolution, the only way that this side-channel-based secure random number generation heuristic will fail is when our technology is able to do two magical things:
1. Create perfectly uniform transistors.
2. Fabricate a complex processor that contains billions of these perfectly uniform transistors without introducing any manufacturing variations (i.e., a perfectly uniform, flawless, manufacturing process)

Additionally, CPUs would have to be designed so that performance is not dynamic (no more temperature, current, voltage, and workload-based dynamic frequency scaling).

Until we reach this level of technology, which does not seem to be on the horizon, and CPUs somehow revert back to having non-dynamic performance scaling features, the proposed side-channel-based heuristic is likely to remain a good candidate for ubiquitous secure random number generation across all our CPU-powered devices.

## 5.0 POTENTIAL IMPACT, USE CASES AND LIMITATIONS

Primarily, I envision SideRand as an open and auditable way for operating system kernels or hypervisors to seed their RNG. Initial seeding during OS installation or first boot is a problem that still needs solving [10], which has led to the vulnerabilities that Heninger et al have discovered [19]. While SideRand is easily extendable to be a general purpose CSPRNG, I'm not very concerned about that right now; there are already great CSPRNGs around, we just need to solve the problem of initial seeding – creating that first 256-bits of entropy in order for all these CSPRNGs to do their job properly. Being open and auditable also means that the very core of our cryptographic security – initial seeding and random number generation – is strengthened against potential backdoor attempts by nation-states.

For servers, SideRand can serve as a replacement for TRNGs. This greatly improves auditability, as SideRand is easily auditable (in its current implementation, a very simple Python 3 source file), whereas TRNGs are impossible to audit in live environments. This will also serve to improve reliability. Hardware eventually fails, even known good hardware, whereas known good software does not, and software is far easier to patch in live environments compared to swapping out hardware in case of eventually discovered defects or needed tuning.

One run of SideRand can replace or complement the entropy collector and random seeder in servers, devices and appliances that would otherwise have poor entropy, solving the problems encountered by Heninger et al [19]. A single run of SideRand can produce a key, from which an unlimited number of keys can then be derived using standard deterministic cryptography, such as using the SideRand-produced key as the nonce or initialization vector to a block cipher in counter mode (e.g. AES-256-CTR) [9].

Headless or embedded devices that previously suffer from "boot-time entropy-hole" can run SideRand during boot (or first-boot only, depending on whether the entropy hole occurs every boot, or only during the first boot) and can afford for SideRand to run for a minute or so (depending on the available timer resolution) to produce the needed strong key, and from there generate secure random numbers using traditional cryptography, such as the aforementioned construction of a secure block cipher in counter mode.

Since target devices may range from large servers to small devices with micro-controllers like Arduino or RaspberryPi, tuning issues will matter. SideRand can be tuned to handle faster processors or low-res timers by increasing the scale factor. Increasing the number of samples collected (which default to 100) may also be a useful tuning parameter.

Processors or platforms with very low-res timers may run into a practical, usability limit – for example, needing several minutes to produce a secure random number at first boot, which may not be acceptable to device manufacturers or vendors. This was simulated in an experiment in Windows, using Python 3 *time.time()* which only has a timer resolution of 16ms. Even increasing the *scale* variable to 10 million and taking 96 seconds to complete one run of SideRand, the quality of the collected values were abysmal – only 151 unique values out of 31,500 measurements (less than half a percent was unique), and the distribution was extremely biased towards a few values. This may be improved with a generous increase in *scale*, but since this already needs 96 seconds to run, increasing *scale* by a factor of 10 or more is already impractical. With a timer resolution of only milliseconds, SideRand is likely to be impractical, collect significantly weaker entropy, and not recommended for use. Data for this very-low-res-timer experiment is also available as supplemental information in the SideRand website. Increasing the number of samples collected is likely to solve this, but further experiments are needed to test this scenario.

## 6.0 CONCLUSION

I presented SideRand, a heuristic and prototype for generating secure random numbers based on the inherent variability of benchmark runtimes, with its worst-case entropy estimate shown to far exceed the required entropy in order to be considered cryptographically secure.

I also proposed two criteria for confidence in sources of secure random numbers, *openness* and *auditability*, and showed how the SideRand heuristic and prototype meets them. The design evolution of the SideRand prototype, from mark 1 to mark 10, was also presented in order to better illustrate the design decisions taken in SideRand and the rationale behind them.

The SideRand mk10 prototype analyzed in this paper, however, is far from widely production-ready. Being a Python 3 prototype, it is only readily deployable in platforms with Python 3 support. The *scale* factor may also need tuning for different platforms.

Further research and development work is recommended in the following areas:
1. Automatic tuning – a separate component should be developed that sets the *scale* factor dynamically. It is recommended that this is a distinct component with its own source code separate from the SideRand source itself, so as to minimize negatively affecting the auditability of the code. Automatic tuning could work as follows: during first run, the tuning component runs SideRand with default scale, measures the unique values being gathered similar to the unique_logger tool, then increases/decreases/does nothing to the scale value depending on the measured values. The ideal values and adjustments would be part of this further research.
2. More platforms – only Intel and AMD processors have been tested, from small-core architectures (Intel Atom, AMD Bobcat) to large-core systems (Haswell & Skylake architectures, Pentium and i7 processors). While SideRand does not depend, in theory and in implementation, on any specific CPU instruction, further tests on other platforms is recommended – for example, micro-controllers like Arduino and Raspberry Pi. Different virtualized environments should also be tested.

3. More languages / software stacks – while Python 3 is a good choice for prototyping and is easily deployable in most x86 servers, be they Windows or Linux, it isn't the most convenient choice for smaller platforms like network devices or appliances and embedded systems. Implementing and testing the SideRand heuristic in a more micro-controller-friendly language or development platform is recommended.
4. More detailed analysis of the entropy in the system – as currently implemented, SideRand's worst-case entropy estimate exceeds what is needed for cryptographic use. While this is good for most cases, for platforms that may encounter usability limits due to having a very slow processor coupled with a very low-res timer, running an overkill algorithm does not help. Further studies and analyses that refine the entropy estimate such that the effective workload can be significantly lessened (e.g., less scale, less samples) may make SideRand more accessible to these devices with slow processors and low-res timers.

I hope that the resulting community review, scrutiny and acceptance of the SideRand heuristic, prototype, and derivatives will help prevent vulnerabilities and other cryptography fails that stem from the difficulty of generating secure random numbers.